\documentclass[twocolumn,eqsecnum,aps,prb]{revtex4}% Physical Review B

\usepackage{graphicx}%
\usepackage{dcolumn}
\usepackage{amsmath}

\makeatletter
\def\btt#1{\texttt{\@backslashchar#1}}%
\DeclareRobustCommand\bblash{\btt{\@backslashchar}}%
\makeatother

\begin{document}

\title{Transport properties of the heavy fermion superconductor PrOs$_{4}$Sb$_{12}$}

\author{H.~Sugawara,\cite{Tokushima} M.~Kobayashi, S.~Osaki, S.R.~Saha, T.~Namiki, Y.~Aoki, and H.~Sato}

\affiliation{Graduate School of Science, Tokyo Metropolitan University, Minami-Ohsawa, Hachioji, Tokyo 192-0397, Japan}

\date{\today}

\begin{abstract}
We have measured the electrical resistivity, thermoelectric power, Hall coefficient, and magnetoresistance (MR) on single crystals of PrOs$_{4}$Sb$_{12}$, LaOs$_{4}$Sb$_{12}$ and NdOs$_{4}$Sb$_{12}$. All the transport properties in PrOs$_{4}$Sb$_{12}$ are similar to those in LaOs$_{4}$Sb$_{12}$ and NdOs$_{4}$Sb$_{12}$ at high temperatures, indicating the localized character of 4$f$-electrons. The transverse MR both in LaOs$_{4}$Sb$_{12}$ and PrOs$_{4}$Sb$_{12}$ tends to saturate for wide field directions, indicating these compounds to be uncompensated metals with no open orbit. We have determined the phase diagram of the field induced ordered phase by the MR measurement for all the principle field directions, which indicates an unambiguous evidence for the $\Gamma _{\rm 1}$ singlet crystalline electric field ground state. \\
\end{abstract}

%\pacs{74.70.Tx, 72.15.Ev, 75.47.-m, 71.27.+a}

\maketitle
\section{INTRODUCTION}
PrOs$_{4}$Sb$_{12}$ with the filled skutterudite structure ($Im$\={3}) was reported to be the first Pr-based heavy fermion (HF) superconductor.~\cite{Bauer,Maple} The heavy mass has been suggested by the large specific heat jump $\Delta C/T_{\rm C}$$\sim$500~mJ/K$^{2}\cdot$ mol at the superconducting (SC) critical temperature $T_{\rm C}=$1.85~K, and directly confirmed by the de Haas-van Alphen (dHvA) experiments.~\cite{Sugawara}  From the various microscopic measurements; i.e. Sb-nuclear quadrupole resonance (NQR) and muon spin relaxation ($\mu$SR) etc.,~\cite{Kotegawa,MacLaughlin,Aoki_mSR} this material has been recognized as an unconventional superconductor which differs from Ce- and U-based HF-superconductors reported to date. Specific heat and magnetic susceptibility measurements in PrOs$_{4}$Sb$_{12}$ indicate the nonmagnetic crystal electric field (CEF) ground state of Pr$^{3+}$ ions,~\cite{Bauer,Maple,Aoki,Vollmer,Tenya,Tayama} which is a sharp contrast with the existing HF superconductors; the existing HF superconductor has magnetic ground state and the magnetic fluctuation is believed to mediate the superconducting pairing. The unconventional nature of this material is also inferred from the unusual SC multiple phase diagram evidenced by the double SC transitions in the specific heat and also by the thermal conductivity measurements which indicates anisotropic SC gap.~\cite{Maple,Vollmer,Measson,Izawa,Chia} Recent zero-field $\mu$SR measurements have revealed the appearance of spontaneous internal fields below $T_{\rm C}$, providing an evidence for the breaking of time reversal symmetry in the SC state.~\cite{Aoki_mSR} As another interesting feature of this material,  an anomalous field induced ordered phase (FIOP) was first observed in the magnetoresistance (MR) measurement~\cite{Ho_proc} and confirmed by the specific heat measurement under magnetic fields.~\cite{Aoki} Recent neutron diffraction experiments has suggested this ordered phase to be an anitiferro-quadrupolar (AFQ) ordering.~\cite{Kohgi} For better understanding of this exotic SC state, it is quite important to clarify the nature of the field induced ordered state.

The exotic superconducting properties of PrOs$_4$Sb$_{12}$ have promoted intense research activities based on various experimental techniques mentioned above. In contrast, the physical properties of the reference compounds LaOs$_4$Sb$_{12}$ and NdOs$_4$Sb$_{12}$ have been only poorly investigated despite its importance for understanding the unusual properties of PrOs$_4$Sb$_{12}$. For the transport properties of PrOs$_4$Sb$_{12}$, the reported data until now have been limited mostly to the electrical resistivity $\rho(T)$ and MR.~\cite{Maple,Ho_proc,Sugawara_LT23,Ho,Maple_JPCS} 
Ho {\it et al.} have reported the MR and an anomaly related to FIOP in PrOs$_4$Sb$_{12}$,~\cite{Ho_proc,Ho, Maple_JPCS} although their measurement is limited only on MR without any description on the angular dependence. Frederick and Maple analyzed their $\rho(T)$ and MR data based on the CEF theory and suggested that the CEF ground state to be $\Gamma_{3}$ ($\Gamma_{\rm 23}$ in $T_{\rm h}$ notation~\cite{Takegahara}) doublet.~\cite{Frederick} Their interpretation of CEF as
an origin of both the decrease (or roll-off) in $\rho(T)$ below 5 K and the FIOP may be correct, however, recent experiments show inconsistency with the $\Gamma_{3}$ ground state model; the results of recent specific heat, magnetization and neutron experiments,~\cite{Aoki,Tenya,Tayama,Kohgi,Kuwahara,Goremychkin} are well describable by the $\Gamma_{1}$ singlet ground state model. Therefore, it is important to investigate anisotropy of FIOP to settle the CEF ground state.

In this paper, we report the systematic study of transport measurements; i.e., electrical resistivity $\rho$, thermoelectric power (TEP) $S$, and Hall coefficient $R_{\rm H}$ in comparison with those on the reference compounds LaOs$_{4}$Sb$_{12}$ and NdOs$_4$Sb$_{12}$. We have also measured the MR for the magnetic fields along all the principle crystalline directions to investigate the anomalous FIOP.

\section{EXPERIMENTAL}
High quality single crystals of $R$Os$_4$Sb$_{12}$ [$R$(rare earth); La, Pr and Nd] were grown by Sb-self-flux method starting from a composition $R$:Os:Sb=1:4:20,~\cite{Takeda,Bauer_CeOs4Sb12}  using high-purity raw materials 4N (99.99\% pure)-La, 4N-Pr, 4N-Nd, 3N-Os and 6N-Sb. The typical forms of the single crystals are cubic or rectangular shape with a largest dimension of about 3 mm. By powder X-ray diffraction experiments, we confirmed that the lattice constants agree with the reported values,~\cite{Braun} and the absence of impurity phases within the experimental accuracy. The residual resistivity ratios (RRR) of the present samples are $\sim 100$ for LaOs$_4$Sb$_{12}$, $\sim 50$ and $\sim 36$ for PrOs$_4$Sb$_{12}$(\#1) and PrOs$_4$Sb$_{12}$(\#2), respectively, indicating the high quality of the present samples, as was confirmed by the observation of the dHvA oscillations.~\cite{Sugawara} The quality of NdOs$_4$Sb$_{12}$ is slightly lower (RRR$\sim 18$).
 
$\rho$ and $R_{\rm H}$ were measured by the ordinary four-probe DC method. $S$ was measured by the differential method using Au-0.07\%~Fe versus Chromel thormocouples. 
The high field MR was measured in a top loading $^3$He-refrigerator cooled down to 0.3~K with a 16~T superconducting magnet.

\section{RESULTS AND DISCUSSION}
\subsection{Electrical resistivity}

\begin{figure}[h]
\begin{center}\leavevmode
\includegraphics[width=0.8\linewidth]{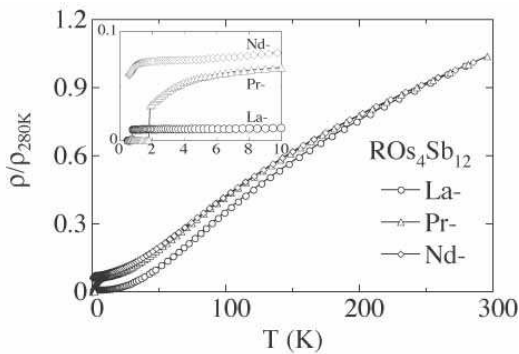}
\caption{Temperature dependence of the electrical resistivity normalized at 280~K in $R$Os$_4$Sb$_{12}$ ($R$; La, Pr and Nd). The inset shows the expanded view below 10~K.}
\label{R}
\end{center}
\end{figure}

Figure~\ref{R} shows the temperature dependence of electrical resistivity $\rho(T)$ normalized at 280~K for $R$Os$_4$Sb$_{12}$ ($R$; La, Pr and Nd).~\cite{R_nor} $\rho(T)$ for PrOs$_4$Sb$_{12}$ is qualitatively the same as the previous reports.~\cite{Bauer,Maple} The similar metallic behavior of $\rho(T)$ for these compounds above $\sim10$~K suggests a localized nature of 4$f$-electrons in $R$Os$_4$Sb$_{12}$ ($R$; Pr and Nd).
As shown in the inset of Fig.~1, we have found the superconductivity in LaOs$_4$Sb$_{12}$ below 0.74~K, which was also confirmed by the NQR and specific heat measurements.~\cite{Kotegawa,Aoki_super} It should be noted that the $T_{\rm C}$ 
of PrOs$_4$Sb$_{12}$ is higher than that of LaOs$_4$Sb$_{12}$. That is 
unusual, since PrOs$_4$Sb$_{12}$ contains Pr-ions with 4$f$-electrons. Actually, for the ordinary Pr-based filled skutterudite superconductor PrRu$_4$Sb$_{12}$ ($T_{\rm C}\sim$1K),  $T_{\rm C}$ is lower than that for LaRu$_4$Sb$_{12}$ ($T_{\rm C}\sim$3.5K).~\cite{Takeda_JPSJ,AbeK}
%%%%%%%%%%%%%%%%%%%%%%%%
%%%%%%%%%%%%%%%%%%%%%%%%%

\begin{figure}[h]
\begin{center}\leavevmode
\includegraphics[width=0.85\linewidth]{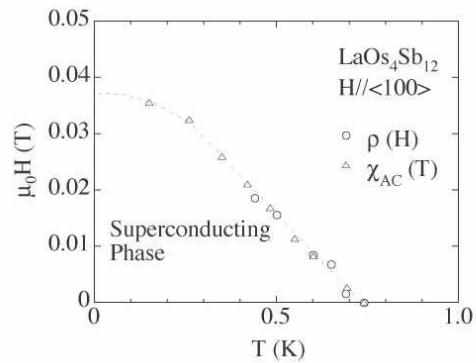}
\caption{Superconducting phase diagram in LaOs$_4$Sb$_{12}$, determined by the field dependence of electrical resistivity and AC susceptibility. The dotted line is guide for eyes.}
\label{LOS_PhaseDiagram}
\end{center}
\end{figure}

Figure~\ref{LOS_PhaseDiagram} shows the SC $H-T$ phase diagram for LaOs$_4$Sb$_{12}$ determined by the field dependence of electrical resistivity at selected temperatures and the temperature dependence of AC susceptibility $\chi_{\rm AC}$ under selected magnetic fields. Maple {\it et al}. also reported the superconductivity of  LaOs$_4$Sb$_{12}$,~\cite{Maple_JPCS} although $T_{\rm C}\sim 1$~K is slightly higher and $H_{\rm C2}\sim 0.6$~T is more than an order of magnitude larger than
that of our results in their experiment. At the present stage, we do not know the origin of such a large discrepancy, though we can say that the sample dependence is not so large within the samples we have grown; we have confirmed the almost the same value of $T_{\rm C}$ and $H_{\rm C2}$ for different several samples and also by different measurements (i.e., $\rho$, $\chi_{\rm AC}$, NQR and specific heat measurement). From the slope of the upper critical field $H_{\rm C2}$ near $T_{\rm C}$, $(-dH_{\rm C2}/dT)_{T_{\rm C}}=$0.068 T/K, the critical field at 0~K $H_{\rm C2}(0)$ and coherence length $\xi_{\rm 0}$ are obtained following Ref.~1, which are summarize in Table~\ref{table1} compared with PrOs$_4$Sb$_{12}$.
%%%%%%%%%%%%%%%%%%%%%%%%
%%%%%%%%%%%%%%%%%%%%%%%%%

%%%%%%%%%%%%%%%%%%%%%%%%%%%%%%%%%%%%%%%%%%%%%%%%%%%
\begin{table}
\caption{Comparison of the superconducting critical temperature $T_{\rm C}$, upper critical field $H_{\rm C2}$(0), coherence length $\xi_{\rm 0}$, penetration depth $\lambda$ and Ginzburg-Landau parameter $\kappa$ between LaOs$_{4}$Sb$_{12}$ and PrOs$_{4}$Sb$_{12}$.}
\centering
\begin{tabular}{@{\hspace{\tabcolsep}\extracolsep{\fill}}lcc}
\hline 
\multicolumn{1}{c}{} & \multicolumn{1}{c}{LaOs$_{4}$Sb$_{12}$} & \multicolumn{1}{c}{PrOs$_{4}$Sb$_{12}$} \\
\hline 
     $T_{\rm C}$ (K) & 0.74$^{a}$ & 1.85$^{b}$ \\
     $H_{\rm C2}$(0) (T) & 0.035$^{a}$ & 2.45$^{b}$ \\
     $\xi_{\rm 0}$ (${\rm \AA}$) & 970$^{a}$  & 116$^{b}$  \\
     $\lambda$ (${\rm \AA}$) & 4700$^{d}$  & 3440$^{c}$  \\
     $\kappa=\lambda /\xi_{\rm 0}$ & 4.8  & 30  \\
\hline
$^{a}$Present work\\
$^{b}$Reference~\cite{Bauer}\\
$^{c}$Reference~\cite{MacLaughlin}\\
$^{d}$Reference~\cite{Aoki_super}\\
\end{tabular}
\label{table1}
\end{table}
%%%%%%%%%%%%%%%%%%%%%%%%%%%%%%%%%%%%%%%%%%%%%%%%%%%

The rapid decrease in $\rho(T)$ for NdOs$_{4}$Sb$_{12}$ below $\sim0.8$~K as shown in the inset of Fig.~\ref{R} might indicate a ferromagnetic transition, since it
shifts to higher temperatures with increasing magnetic field.

\subsection{Thermoelectric power}

\begin{figure}[h]
\begin{center}\leavevmode
\includegraphics[width=0.85\linewidth]{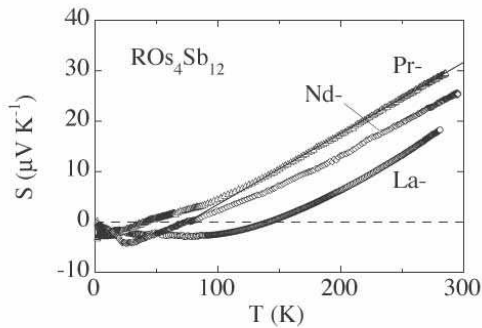}
\caption{Temperature dependence of the thermoelectric power in $R$Os$_4$Sb$_{12}$ ($R$; La, Pr and Nd).}
\label{S}
\end{center}
\end{figure}

Figure.~\ref{S} shows the temperature dependence of TEP $S(T)$ for $R$Os$_{4}$Sb$_{12}$ ($R$; La, Pr and Nd).
%%%%%%%%%%%%%%%%%%%%%%%%
%%%%%%%%%%%%%%%%%%%%%%%%%
The TEPs of all the compounds are not describable only by a simple combination of the diffusion and the phonon drag TEP. For LaOs$_{4}$Sb$_{12}$, the TEP shows a broad negative peak near 80~K, for which there are two possible explanations. The peak temperature is not so far from  $\Theta_{\rm D}/5(\sim60$ K) expected  for the phonon drag peak (one fifths of the Debye temperature $\Theta_{\rm D}$), taking into account the reported value of $\Theta_{\rm D}=304$ K.~\cite{Bauer_CeOs4Sb12}  However, PrOs$_{4}$Sb$_{12}$ does not show such a structure though it has almost the same Debye temperature, that in turn suggests the phone drag is not the main contribution for such a low temperature structure in LaOs$_{4}$Sb$_{12}$. Another possible origin is some fine structure in the density of states $N(\varepsilon)$ near the Fermi level $\varepsilon_{\rm F}$, since the diffusion themoelectric power can be represented as  $-(\pi^{2}k_{B}^{2}T/3|e|)[dN(\varepsilon)/d\varepsilon]_{\varepsilon_{\rm F}}$ in the simplest free electron model.~\cite{Dugdale} 
According to the band structure calculations,~\cite{Harima_LaOs4Sb12} $\varepsilon_{\rm F}$ is located in between a large peak and a small peak in the density of states; the energy difference between these two peaks is $\Delta E/k_{\rm B}\sim$100~K. At low temperature, the positive slope of $N(\varepsilon)$ at $\varepsilon_{\rm F}$ leads to the negative $S(T)$, With increasing temperature, the small peak structure becomes thermally smeared, and the averaged slope of $N(\varepsilon)$ near $\varepsilon_{\rm F}$ becomes negative, reading to the positive $S(T)$ at high temperature. Such temperature dependence of averaged slope of $N(\varepsilon)$ may be an origin of the low temperature structure of TEP in $R$Os$_{4}$Sb$_{12}$. The slope of $S(T)$ at higher temperatures is not much different among the three compounds, indicating basically the same electronic structure. The slope $dS/dT\sim 0.14\mu$V/K$^{2}$ is an order of magnitude large compared to ordinary $sp$-metals and is similar to 3$d$-transition metals such as Ni, Pd, and Pt,~\cite{Barnard} that is consistent with the large electronic density of states at Fermi level predicted by the band structure calculation.~\cite{Sugawara,Harima_LaOs4Sb12}

\subsection{Hall coefficient}

\begin{figure}[h]
\begin{center}\leavevmode
\includegraphics[width=0.85\linewidth]{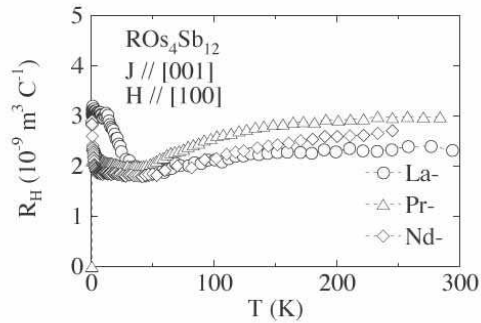}
\caption{Temperature dependence of the Hall coefficient in $R$Os$_4$Sb$_{12}$ ($R$; La, Pr and Nd).}
\label{RH}
\end{center}
\end{figure}

Figure.~\ref{RH} shows the temperature dependence of Hall coefficient $R_{\rm H}(T)$ for $R$Os$_{4}$Sb$_{12}$ ($R$; La, Pr and Nd).
%%%%%%%%%%%%%%%%%%%%%%%%
%%%%%%%%%%%%%%%%%%%%%%%%%
The Hall coefficient is almost temperature independent at high temperatures irrespective of $R$. The absolute value is also not much different among the three compounds, taking into account relatively large error in the geometrical determination. The carrier number at higher temperatures for PrOs$_4$Sb$_{12}$ is $n=2.1\times10^{27}/{\rm m}^{3}$ (0.85-holes/molecular-unit) assuming a single carrier model. That is consistent with the result of dHvA experiments,~\cite{Sugawara} which is well explained by the band structure calculation predicting two pieces of hole-like Fermi surfaces (FS) and a multiply connected one; the number of holes within the two hole-like surfaces gives $n=1.1\times10^{27}/{\rm m}^{3}$. The weak temperature dependence below $\sim150$~K could be ascribed to the temperature dependent anisotropy of relaxation time,~\cite{Tsuji} which is not unusual as was observed even in simple metals such as Al and Pb.~\cite{Sato2} For LaOs$_4$Sb$_{12}$, the decrease down to 40~K is ascribed to the change in the main scattering centers from the isotropic phonon-scattering with large wave vectors ${\bf q}$ to the anisotropic phonon-scattering with smaller ${\bf q}$, and the increase below $\sim40$~K reflects the recovery to the isotropic scattering by impurities. The rapid changes both in PrOs$_4$Sb$_{12}$ and NdOs$_4$Sb$_{12}$ at low temperature are ascribed to the superconducting and ferromagnetic transition, respectively.

On $\rho(T)$, $S(T)$, and $R_{\rm H}(T)$ in PrOs$_4$Sb$_{12}$, qualitative comparison to those in PrFe$_4$P$_{12}$ might be of interest, since both compounds exhibit HF behaviors which are very rare as Pr-based compounds. Among Ce-based dense Kondo compounds, -ln$T$ dependence in  $\rho(T)$, a large positive peak in $S(T)$, anomalous Hall effect in $R_{\rm H}(T)$ are well known as common features associated with the Kondo scattering (cooperated with the crystal field).  For PrFe$_4$P$_{12}$, an apparent -ln$T$ dependence in $\rho(T)$, huge negative TEP peak ($\sim 70\mu$V/K), and the skew scattering in $R_{\rm H}(T)$ have been found above the AFQ transition temperature,~\cite{Sato} which are thought to be supportive evidences of an intense Kondo effect. In contrast, no such drastic features appear in PrOs$_4$Sb$_{12}$ indicative well localized nature of 4$f$-electrons. 
For PrFe$_4$P$_{12}$ wtih the smaller lattice constant ($\simeq7.8$\AA), the stronger $c-f$ hybridization is expected. On the other hand, for PrOs$_4$Sb$_{12}$ with larger lattice constant ($\simeq9.3$\AA), the $c-f$ hybridization is expected to be weaker. The difference of $c-f$ hybridization strength might be the origin of large differences of transport properties between these two compounds.
 
\subsection{Magnetoresistance}

\begin{figure}[h]
\begin{center}\leavevmode
\includegraphics[width=0.8\linewidth]{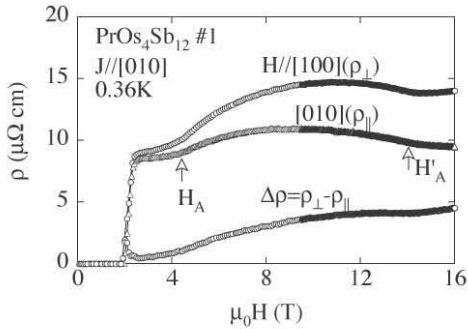}
\caption{Comparison of the field dependence of magnetoresistance both for the transverse ($\rho_{\rm \perp}$) and longitudinal ($\rho_{\rm \parallel}$) geometry in PrOs$_4$Sb$_{12}$.}
\label{MR_tate_yoko}
\end{center}
\end{figure}

\begin{figure}[t]
\begin{center}\leavevmode
\includegraphics[width=0.8\linewidth]{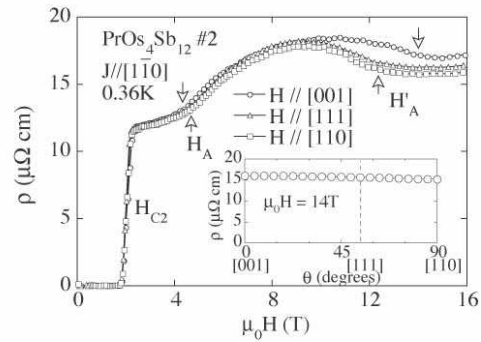}
\caption{Field dependences of the transverse magnetoresistance for field along the all principal directions in PrOs$_4$Sb$_{12}$. The inset shows the angular dependence of the transverse magnetoresistance.}
\label{MR_A_H}
\end{center}
\end{figure}

\begin{figure}[t]
\begin{center}\leavevmode
\includegraphics[width=0.8\linewidth]{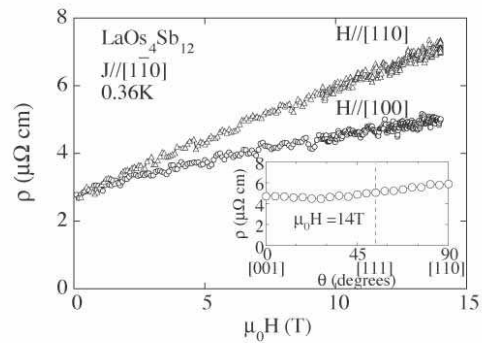}
\caption{Field dependences of the transverse magnetoresistance for field along the [100] and [110] directions in LaOs$_4$Sb$_{12}$. The inset shows the angular dependence of the transverse magnetoresistance.}
\label{LOS_MR_A_H}
\end{center}
\end{figure}

Figure~\ref{MR_tate_yoko} shows the comparison of the field dependence of MR between the transverse ($\rho_{\rm \perp}$) and longitudinal ($\rho_{\rm \parallel}$) geometry in PrOs$_4$Sb$_{12}$.
%%%%%%%%%%%%%%%%%%%%%%%%
%%%%%%%%%%%%%%%%%%%%%%%%%
The difference between the two geometries, $\Delta \rho=\rho_{\rm \perp}-\rho_{\rm \parallel}$, is positive and could be mainly ascribed to the ordinary Lorentz MR contribution. Figures~\ref{MR_A_H} and \ref{LOS_MR_A_H} show the angular and field dependences of transverse MR for the field along principal crystallographic directions in PrOs$_4$Sb$_{12}$ and LaOs$_4$Sb$_{12}$, respectively, where $\theta$ both in the insets of Figs.~\ref{MR_A_H} and \ref{LOS_MR_A_H} represents to the field angle from the [001] direction.
%%%%%%%%%%%%%%%%%%%%%%%%
%%%%%%%%%%%%%%%%%%%%%%%%%
%%%%%%%%%%%%%%%%%%%%%%%%
%%%%%%%%%%%%%%%%%%%%%%%%%
As shown in Fig.~\ref{MR_tate_yoko}, $\Delta \rho$ shows a saturating tendency with increasing field. Combined with the small angular dependence in the inset of Fig.~\ref{MR_A_H}, PrOs$_4$Sb$_{12}$ is judged to be an uncompensated metal with no open orbit. That is also confirmed in LaOs$_4$Sb$_{12}$ as shown in Fig.~\ref{LOS_MR_A_H}, whose FS topology is almost the same as that in PrOs$_4$Sb$_{12}$.~\cite{Sugawara}  The difference in angular dependence between those insets of  Figs.~\ref{MR_A_H} and \ref{LOS_MR_A_H} may be ascribed to 4$f$-contribution in PrOs$_4$Sb$_{12}$ where the larger residual resistivity suppresses the ordinary MR contribution compared to LaOs$_4$Sb$_{12}$.

As shown in Fig.~\ref{MR_tate_yoko}, for both the transverse and longitudinal geometries in PrOs$_4$Sb$_{12}$, there exists positive contribution due to the scattering with magnetic and/or orbital degree of freedom, which is dominating in the field induced ordered phase inferred from the two anomalies, $H_{\rm A}$ and $H'_{\rm A}$, indicated by the arrows. Important finding in this experiments is the existence of the anomaly in MR indicating the phase boundary of the FIOP  for all the principal field directions as shown in Fig.~\ref{MR_A_H}.
The result for $H\|$[100] agrees with that reported by Ho {\it et al.} except the absolute value of the resistivity.~\cite{Ho} The dome-like structure between $H_{\rm A}$ and $H'_{\rm A}$ has been explained by Frederick and Maple as due to the conduction electron scattering associated with the CEF excitation.~\cite{Frederick}  
Taking into account the fact that the magnetic moment of Pr is enhanced in the FIOP,~\cite{Kohgi} the resistivity increase in the FIOP could be most naturally ascribed to the enhancement of electron scattering associated with the CEF excitation. The first CEF excited state is reported to locate above $\Delta_{\rm 1}/k_{\rm B}\sim 8$~K both in $\Gamma_{\rm 3}$ and $\Gamma_{\rm 1}$ ground state models,~\cite{Bauer,Aoki,Tayama,Kohgi} which naturally explains the gradual decrease of $\rho(T)$ below $\sim$5~K under 0~T as shown in the inset of Fig.~\ref{R}. At low temperatures far below $\Delta_{\rm 1}/k_{\rm B}$, the electron scattering associated with CEF excitation is suppressed in zero field. For the $\Gamma_{\rm 1}$ ground state model,~\cite{Aoki,Tayama,Kohgi} with increasing field, one out of $\Gamma_{\rm 5}$($\Gamma_{\rm 4}^{\rm (2)}$ in $T_{\rm h}$ notation~\cite{Takegahara}) triplet excited state comes down close to the ground state, resulting in enhancement of electrical resistivity associated with CEF excitation. After showing a maximum at a crossing field $\sim 9$~T of the two levels, the electrical resistivity decreases with increasing gap between the two levels. The closeness of the two levels is  thought to be a key factor to stabilize the FIOP.~\cite{Kohgi} It should be noted that the level crossing is expected for field along the all principal crystallographic directions in the $\Gamma_{\rm 1}$ ground state model,~\cite{Tayama,Shiina1,Shiina2} whereas no level crossing is expected for fields parallel to the $\langle 110\rangle$ and $\langle 111\rangle$ directions in the $\Gamma_{\rm 3}$ ground state model.~\cite{Tayama} Thus the present result indicates an unambiguous evidence for the $\Gamma _{\rm 1}$ CEF ground state in PrOs$_4$Sb$_{12}$.

\begin{figure}[!tbp]
\begin{center}\leavevmode
\includegraphics[width=0.75\linewidth]{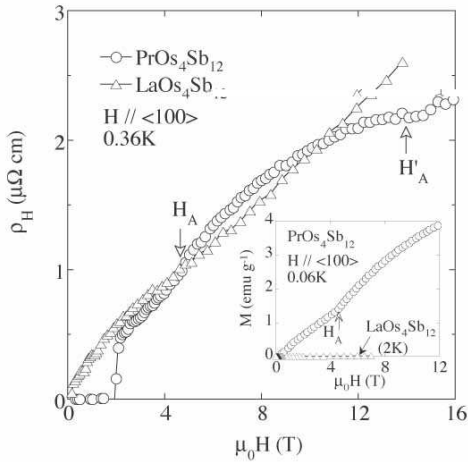}
\caption{Comparison of the field dependence of Hall resistivity between PrOs$_4$Sb$_{12}$ and LaOs$_4$Sb$_{12}$. The inset shows the comparison of magnetization curves between these two compounds.}
\label{RH_H}
\end{center}
\end{figure}

As Frederick and Maple discussed,~\cite{Frederick} the dome-like structure in $\rho(H)$ near 9~T at low temperatures may be ascribed to the CEF excitation. In addition, we discuss here the origin of the sharp anomaly across the phase boundary. One possiblity is the slight change of FS associated with the super-zone gap formation of new periodicity; namely the AFQ ordering as was clarified by the neutron scattering experiment.~\cite{Kohgi} Actually, as shown in Fig.~\ref{RH_H}, the field dependence of Hall resistivity $\rho_{\rm H}(H)$ exhibits the slight slope changes across $H_{\rm A}$ and $H'_{\rm A}$, in contrast with the almost linear
increase of $\rho_{\rm H}(H)$ in LaOs$_4$Sb$_{12}$.
%%%%%%%%%%%%%%%%%%%%%%%%
%%%%%%%%%%%%%%%%%%%%%%%%%
Also in the dHvA experiments,~\cite{Sugawara} the frequency of  $\beta$-branch originating from the 48th hole FS has been confirmed to exhibit slight changes across $H_{\rm A}$ and $H'_{\rm A}$. However, it could not be a decisive evidence for the FS change because the magnetization also changes non-linearly at $H_{\rm A}$ as shown in the inset of Fig.~\ref{RH_H},~\cite{Tenya,Tayama} leading to a superficial change in FS. Such a slight change of the dHvA frequency can be also ascribed to the combined effect of the spin splitting and non-linear magnetization.~\cite{Sugawara_PRB} 
Thus, the change in FS across the FIOP boundary has not been settled yet, however, taking into account the normal Hall contribution is dominating in the Hall resistivity as shown in Fig~\ref{RH_H}; namely the Hall resistivities of two compounds with the same sign and similar magnitude naturally suggest the anomalous Hall effect~\cite{Fert} is not dominant in PrOs$_4$Sb$_{12}$, the FS does not drastically change in FIOP, that is in contrast with PrFe$_4$P$_{12}$ in which the apparent FS reconstruction has been observed across the AFQ ordering.~\cite{Sugawara_PRB}

\begin{figure}[!tbp]
\begin{center}\leavevmode
\includegraphics[width=1\linewidth]{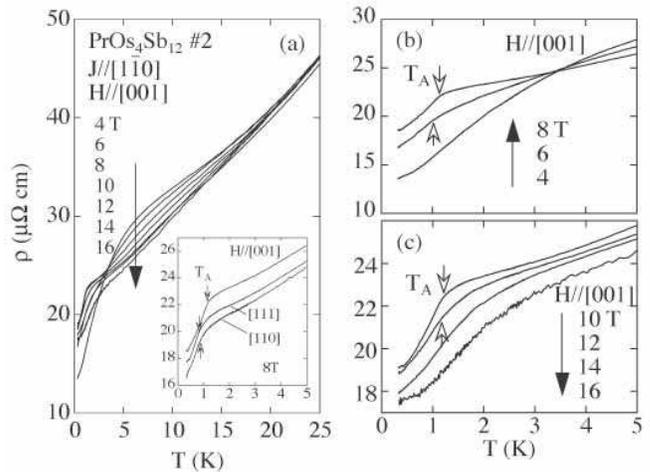}
\caption{Temperature dependence of the resistivity $\rho(T)$ at selected magnetic fields in PrOs$_4$Sb$_{12}$. The inset in Fig.9(a) shows anisotropy of $\rho(T)$ at 8 T.  Fig.9(b) and Fig.9(c) show $\rho(T)$ below  8 T and above 10 T, respectively. $T_{\rm A}$ represents to the transition temperature of FIOP.}
\label{MR_T}
\end{center}
\end{figure}

\begin{figure}[!tbp]
\begin{center}\leavevmode
\includegraphics[width=0.8\linewidth]{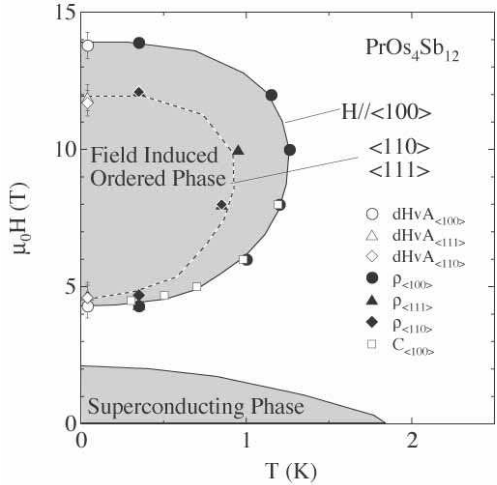}
\caption{$H-T$ phase diagram of PrOs$_4$Sb$_{12}$.}
\label{PhaseDiagram}
\end{center}
\end{figure}

In order to determine the $H-T$ phase diagram for field along the three principal field directions, we have measured $\rho(T)$ for selected fields as shown in Fig.~\ref{MR_T}. 
%%%%%%%%%%%%%%%%%%%%%%%%
%%%%%%%%%%%%%%%%%%%%%%%%%
We have also measured $\rho(T)$ for $H\|[110]$ and $H\|[111]$, which are basically the same as that for $H\|[100]$ except the difference in transition temperature $T_{\rm A}$ as shown in the inset of Fig.~\ref{MR_T} (a). A shoulder (or roll-off) of $\rho(T)$ around $\sim5$~K associated with the CEF excitation is strongly suppressed by magnetic fields as shown in Fig.~\ref{MR_T} (a), suggesting the CEF excited state comes down approaching the ground state. As shown in Figs.~\ref{MR_T} (b) and \ref{MR_T} (c), above 6~T, we can see a bend below $T_{\rm A}=1.25$~K, which is apparently related with FIOP. With increasing magnetic fields, $T_{\rm A}$ increases and the bend becomes sharper (i.e., at 8~T and 10~T). After showing a maximum around 10~T, $T_{\rm A}$ decreases rapidly and the anomaly disappears above 14~T. 
Figure~\ref{PhaseDiagram} shows the $H-T$ phase diagram determined by the present MR measurements along with specific heat and dHvA experiments. 
%%%%%%%%%%%%%%%%%%%%%%%%
%%%%%%%%%%%%%%%%%%%%%%%%%
The similar phase diagram has already reported by other groups,~\cite{Ho,Maple_JPCS,Oeschler} however, there are no description about the field anisotropy in their reports. In the anisotropy determination of FIOP based on the magnetization measurements,~\cite{Tayama} the field range was limited to only below 13 T. In the present MR measurements up to 16~T, the upper boundary $H'_{\rm A}$ of FIOP including the anisotropy has been determined for all the principal field directions for the first time; $H'_{\rm A}\sim14$~T for $H\|\langle100\rangle$ and $H'_{\rm A}\sim12$~T for $H\|\langle110\rangle$ and $H\|\langle111\rangle$. We again emphasize that the FIOP exists for the field along all the three principal crystallographic directions.
The presence of  AFQ phase close proximity to the superconducting phase remind us the Ce- and U-based HF-superconductors and high-$T_{\rm C}$ oxides; in those systems the Cooper pairing is believed to be mediated by magnetic fluctuations.~\cite{Mathur} By analogy, in PrOs$_{4}$Sb$_{12}$, the quadrupole fluctuations of Pr-ions might play an important role in the HF-SC properties.
 
\section{CONCLUSIONS}
All the transport properties of PrOs$_{4}$Sb$_{12}$, compared to those of the references LaOs$_{4}$Sb$_{12}$ and NdOs$_{4}$Sb$_{12}$, suggest the 4$f$-electrons to be well localized at high temperatures, while the exotic behaviors associated with 4$f$-electrons degree of freedom are observed below the temperature of order of CEF splitting between the ground state and the first excited state.

From the field dependence of MR both for the transverse and longitudinal geometry in PrOs$_4$Sb$_{12}$, we found that the ordinary MR contribution saturates for all the field directions. This fact suggests that PrOs$_4$Sb$_{12}$ is an uncompensated metal with no open orbit, which is consistent with the band structure calculation.

The dependences of MR on both field and temperature have revealed that the FIOP exists for the field along all the three principal crystallographic directions between 4.4~T and 14~T below 1.25~K, which strongly supports the $\Gamma_{\rm 1}$ CEF ground state model proposed based on the specific heat, magnetization and neutron diffraction measurements.

\begin{acknowledgments}
We thank Professor H.~Harima, Professor M.B.~Maple and Professor K.~Miyake for helpful discussion. This work was supported by a Grant-in-Aid for Scientific Research Priority Area "Skutterudite" (No.15072206) of the Ministry of Education, Culture, Sports, Science and Technology, Japan.

\end{acknowledgments}


\begin{thebibliography}{}


\bibitem[*]{Tokushima}Present address: Department of Mathematical and Natural Sciences, Faculty of Integrated Arts and Sciences,The University of Tokushima, Tokushima, 770-8502, Japan, E-mail:sugawara@ias.tokushima-u.ac.jp
%\bibitem[**]{KEK}Present address: Meson Science Laboratory, Institute of Materials Structure Science, KEK, Tsukuba, Ibaraki 305-0801, Japan.
%\bibitem[***]{Osaka}Present address: 
\bibitem{Bauer} E.D.~Bauer, N.A.~Frederick, P.-C.~Ho, V.S.~Zapf, and M.B.~Maple, Phys. Rev. B {\bf 65}, 100506(R) (2002).
\bibitem{Maple} M.B.~Maple, P.-C.~Ho, V.S.~Zapf, N.A.~Frederick, E.D.~Bauer, W.M.~Yuhasz, F.M.~Woodward, and J.W.~Lynn, J. Phys. Soc. Jpn. Suppl. {\bf 71}, 23 (2002).
\bibitem{Sugawara}H.~Sugawara, S.~Osaki, S.R.~Saha, Y.~Aoki, H.~Sato, Y.~Inada, H.~Shishido, R.~Settai, Y.~\={O}nuki, H.~Harima, and K.~Oikawa, Phys. Rev. B {\bf 66}, 220504(R) (2002).
\bibitem{Kotegawa}H.~Kotegawa, M.~Yogi, Y.~Imamura, Y.~Kawasaki, G.-q. Zheng, Y.~Kitaoka, S,~Ohsaki, H.~Sugawara, Y.~Aoki, and H.~Sato, Phys. Rev. Lett {\bf 90}, 027001 (2003).
\bibitem{MacLaughlin}D.E.~MacLaughlin, J.E.~Sonier, R.H.~Heffner, O.O.~Bernal, Ben-Li Young, M.S.~Rose, G.D.~Morris, E.D.~Bauer, T.D.~Do, and M.B.~Maple, Phys. Rev. Lett. {\bf 89}, 157001 (2002).
\bibitem{Aoki_mSR}Y.~Aoki, A.~Tsuchiya, T.~Kanayama, S.R.~Saha, H.~Sugawara, H.~Sato, W.~Higemoto, A.~Koda, K.~Ohishi, K.~Nishiyama and R,~Kadono, Phys. Rev. Lett. {\bf 91}, 067003 (2003).
\bibitem{Aoki}Y.~Aoki, T.~Namiki, S.~Osaki, S.R.~Saha, H.~Sugawara, H.~Sato, J. Phys. Soc. Jpn. {\bf 71}, 2098 (2002).
\bibitem{Vollmer}R.~Vollmer, A.~Fai{\ss}t, C.~Pfleiderer, H.v.~L\"{o}hneysen, E.D.~Bauer, P.-C.~Ho, V.~Zapf, and M.B.~Maple, Phys. Rev. Lett. {\bf 90}, 057001 (2003).
\bibitem{Tenya}K.~Tenya, N.~Oeschler, P.~Gegenwart, F.~Steglich, N.A.~Frederick, E.D.~Bauer, and M.B.~Maple, Acta Physica Polonica B 34, 995 (2003).
\bibitem{Tayama}T.~Tayama, T.~Sakakibara, H.~Sugawara, Y.~Aoki, and H.~Sato, J. Phys. Soc. Jpn {\bf 72} (2003).
\bibitem{Measson}M.-A. Measson, D. Braithwaite, J. Flouquet, G. Seyfarth, J. P. Brison, E. Lhotel, C. Paulsen, H. Sugawara, and H. Sato, Phys. Rev. B {\bf 70}, 064516 (2004).
\bibitem{Izawa}K.~Izawa, Y.~Nakajima, J.~Goryo, Y.~Matsuda, S.~Osaki, H.~Sugawara, H.~Sato, P.~Thalmeier, and K.~Maki, Phys. Rev. Lett {\bf 90}, 117001 (2003).
\bibitem{Chia}E.E.M. Chia, M.B. Salamon, H. Sugawara, and H. Sato, Phys. Rev. Let. {\bf 91}, 247003 (2003).
\bibitem{Ho_proc} P.-C.~Ho, V.S.~Zapf, E.D.~Bauer, N.A.~Frederick, M.B.~Maple, G. Giester, P. Rogl, St. Berger, Ch. Paul and E. Bauer, in {\it Physical Phenomena at High Magnetic Fields-IV,} edited by G. Boebinger, Z. Fisk, L.P. Gorkov, A. Lacerda and J.R. Schrieffer (World Scientific, Singapore, 2001), pp. 98-103.
\bibitem{Kohgi}M.~Kohgi, K.~Iwasa, M.~Nakajima, N.~Metoki, S.~Araki, N.~Bernhoeft, J.M.~Mignot, A.~Gukasov, H.~Sato, Y.~Aoki, and H.~Sugawara, J. Phys. Soc. Jpn {\bf 72} 1002 (2003).
\bibitem{Sugawara_LT23}H.~Sugawara, S.~Osaki, E.~Kuramochi, M.~Kobayashi, S.R.~Saha, T.~Namiki, Y.~Aoki, and H.~Sato, Physica B {\bf 329-333}, 551 (2003). 
\bibitem{Maple_JPCS} M.B.~Maple, P.-C.~Ho, V.S.~Zapf, N.A.~Frederick, W.M.~Yuhasz, E.D.~Bauer, A.D. Christianson and A.H. Lacerda, J. Phys.: Condens. Matter {\bf 15}, S2071 (2003).
\bibitem{Ho} P.-C.~Ho, N.A.~Frederick, V.S.~Zapf, E.D.~Bauer, T.D. Do,M.B.~Maple, A.D. Christianson and A.H. Lacerda, Phys. Rev. B {\bf 67}, 180508(R) (2003).
\bibitem{Takegahara}K.Takegahara, H. Harima and A. Harima, J. Phys. Soc. Jpn. {\bf 70}, 1190 (2001).
\bibitem{Frederick}N.A. Frederick and M.B. Maple, J. Phys.: Condens. Matter {\bf 15}, 4789 (2003).
\bibitem{Kuwahara}K. Kuwahara, K. Iwasa, M. Kohgi, K. Kaneko, S. Araki, N. Metoki, H. Sugawara, Y. Aoki and H. Sato, J. Phys. Soc. {\bf 73}, 1438 (2004).
\bibitem{Goremychkin}E.A. Goremychkin, R. Osborn, E.D. Bauer, M.B. Maple, N.A. Frederick, W. M. Yuhasz, FM. Woodward, and J.W. Lynn, Phys. Rev. Lett. {\bf 93} 157003 (2004).
\bibitem{Takeda}N.~Takeda and M.~Ishikawa, Physica B {\bf 259-261}, 92 (1999).
\bibitem{Bauer_CeOs4Sb12}E.~D.~Bauer, A.~\'{S}lebarski, E.~J.~Freeman, C.~Sirvent, and M.~B.~Maple, J. Phys.: Condens. Matter {\bf 13}, 4495 (2001).
\bibitem{Braun}D.~J.~Braun and W.~Jeitschko, J. Less-Common Met. {\bf 72}, 147 (1980).
\bibitem{R_nor}The absolute values of the resistivity at room temperature are $290\mu\Omega$cm for LaOs$_4$Sb$_{12}$, $390\mu\Omega$cm for PrOs$_4$Sb$_{12}$(\#1), $450\mu\Omega$cm for PrOs$_4$Sb$_{12}$(\#2), and $250\mu\Omega$cm for NdOs$_4$Sb$_{12}$, respectively, Such variation is probably due to the irregular shapes and/or the presence of microcracks in the samples. 
\bibitem{Aoki_super}Y. Aoki, W. Higemoto, S. Sanada, K. Ohishi, S.R. Saha, A. Koda, K. Nishiyama, R. Kadono, H. Sugawara and H. Sato, Physica B {\bf 359-361}, 895 (2005).
\bibitem{Takeda_JPSJ}N.~Takeda and M.~Ishikawa, J. Phys. Soc. Jpn. {\bf 69}, 868 (2000).
\bibitem{AbeK}K.~Abe, H.~Sato, T.D.~Matsuda, N.~Namiki, H.~Sugawara and Y.~Aoki, J. Phys.: Condens. Matter {\bf 14}, 11757 (2002).
\bibitem{Dugdale}J.S.~Dugdale, {\it The Electrical Properties of Metals and Alloys} (Arnold, London, 1977). 
\bibitem{Harima_LaOs4Sb12}H.~Harima and K.~Takegahara, Physica C {\bf 388-389}, 555 (2003).
\bibitem{Barnard}See, for example, R.D.Barnard, {\it Thermoelectricity in Metals and Alloys} (Taylor \& Francis, London, 1972), p.~170.
\bibitem{Sato}H.~Sato, Y.~Abe, H.~Okada, T.~D.~Matsuda, K.~Abe, H.~Sugawara, and Y.~Aoki, Phys. Rev. B {\bf 62}, 15125 (2000).
\bibitem{Tsuji}M.~Tsuji, J. Phys. Soc. Jpn. {\bf 13}, 979 (1958).
\bibitem{Sato2}H.~Sato, H.~Okimoto, I.~Sakamoto and K.~Yonemitsu, J. Phys. F: Met. Phys. {\bf 15}, 1555 (1985).
%\bibitem{Matsuda} T.D.~Matsuda, K. Abe, F. Watanuki, H. Sugawara, Y. Aoki, H. Sato, Y. Inada, R. Settai, and Y.~\={O}nuki, Physica B {\bf 312-313}, 832 (2002).
\bibitem{Shiina1} R. Shiina and Y. Aoki, J. Phys. Soc. Jpn. {\bf 73}, 541 (2004).
\bibitem{Shiina2} R. Shiina, J. Phys. Soc. Jpn. {\bf 73}, 2257 (2004).
\bibitem{Sugawara_PRB}H.~Sugawara, T.~D.~Matsuda, K.~Abe, Y.~Aoki, H.~Sato, S.~Nojiri, Y.~Inada, R.~Settai, and Y.~\={O}nuki, Phys. Rev. B {\bf 66}, 134411 (2002).
\bibitem{Fert}A.~Fert and P.~Levy, Phys. Rev. B {\bf 26}, 1907 (1987).
\bibitem{Oeschler} N. Oeschler, P, Gegenwart, F. Weickert, I. Zerec, P. Thalmeier, F. Steglich, E.D.~Bauer, N.A.~Frederick, M.B.~Maple, Phys. Rev. B {\bf 69}, 235108 (2004).
\bibitem{Mathur}N.D.~Mathur, F.M. Grosche, S.R.~Julian, I.R. Walker, D.M. Freye, R.K.W.~Haselwimmer, and G.G.~Lonzarich, Nature {\bf 394}, 39 (1998).
\end{thebibliography}
\end{document}